# Pomeranchuk instability of a topological crystal

**Authors:** Md Shafayat Hossain[1]*†, Zahir Muhammad[2]*, Rajibul Islam[3]*, Zi-Jia Cheng[1]*, Yu-Xiao Jiang[1], Maksim Litskevich[1], Tyler A. Cochran[1], Xian P. Yang[1], Byunghoon Kim[1], Fei Xue[3], Ilias E. Perakis[3], Weisheng Zhao[2], Mehdi Kargarian[4]†, Luis Balicas[5], Titus Neupert[6], M. Zahid Hasan[1,7,8]†

**Affiliations:**
[1]Laboratory for Topological Quantum Matter and Advanced Spectroscopy, Department of Physics, Princeton University, Princeton, New Jersey, USA.

[2]Hefei Innovation Research Institute, School of Microelectronics, Beihang University, Hefei 230013, P.R. China.

[3]Department of Physics, University of Alabama at Birmingham, Birmingham, Alabama 35294, USA.

[4]Department of Physics, Sharif University of Technology, Tehran 14588-89694, Iran

[5]National High Magnetic Field Laboratory, Tallahassee, Florida 32310, USA

[6]Department of Physics, University of Zurich, Winterthurerstrasse, Zurich, Switzerland.

[7]Princeton Institute for Science and Technology of Materials, Princeton University, Princeton, NJ, USA.

[8]Lawrence Berkeley National Laboratory, Berkeley, California 94720, USA.

†Corresponding authors, E-mail: mdsh@princeton.edu; kargarian@physics.sharif.edu; mzhasan@princeton.edu.

*These authors contributed equally to this work.

**Abstract:**

Nematic quantum fluids appear in strongly interacting systems and break the rotational symmetry of the crystallographic lattice. In metals, this is connected to a well-known instability of the Fermi liquid–the Pomeranchuk instability. Using scanning tunneling microscopy, we identified this instability in a highly unusual setting: on the surface of an elemental topological metal, arsenic. By directly visualizing the Fermi surface of the surface state via scanning tunneling spectroscopy and photoemission spectroscopy, we find that the Fermi surface gets deformed and becomes elliptical at the energies where the nematic state is present. Known instances of nematic instability typically need van-Hove singularities or multi-orbital physics as drivers. In contrast, the surface states of arsenic are essentially indistinguishable from well-confined isotropic Rashba bands near the Fermi level, rendering our finding the first realization of Pomeranchuk instability of the topological surface state.

**One-sentence summary:**

We observe strong interaction physics on the surface of a topological material demonstrated via the Pomeranchuk instability of the topological surface state for the first time.

**Main text:**

Spontaneous symmetry breaking leads to some of the most intriguing phenomena in physics, such as ferromagnetism, liquid crystalline phases in quantum fluids, and the Higgs mechanism for mass generation [1-3]. The nematic electronic state is an exciting example of broken-symmetry states that stem from electron-electron interactions [4-8]. Nematic phases are characterized by a reduced rotational symmetry compared to that of the underlying crystal lattice and those of Fermi liquids. They are theorized to be driven by distortions of the Fermi surface [9,10]. Back in 1958, Pomeranchuk highlighted that if the Fermi liquid parameter $F_l$ in the angular momentum channel $l$ were less than $-(2l+1)$, the Landau Fermi liquid would become unstable against deformation of the Fermi surface in that channel [11]. When microscopic interactions lead to $F_l < -(2l+1)$, an isotropic or nearly spherical Fermi liquid state transitions into a nematic state with an elliptical Fermi surface. Nematic states have been reported in various strongly correlated systems such as doped Mott insulators [5], cuprates [4,12], iron-based superconductors [7, 8], $Sr_3Ru_2O_7$ at high magnetic fields [13], half-filled higher Landau level states [6, 14, 15], quantum Hall liquids on bismuth surfaces at high magnetic fields [16], topological superconductor candidates [17-21] and Kagome superconductors [22, 23]. However, these examples are not pristine realizations of Pomeranchuk's idea of a Fermi liquid instability as they involve either additional broken symmetries or a non-Fermi-liquid state. Here, we present the discovery of a Pomeranchuk instability-driven nematic order in elemental arsenic (α-As), providing a direct experimental confirmation of this fundamental mechanism.

α-As is a clean electronic system, hosting a topologically connected surface state centered around the Γ-point [24, 25] alongside a conjoined strong and higher-order topological insulator phase [25]. At the Fermi level, the surface state resembles a Rashba-split band of a two-dimensional electron gas. This establishes α-As as a versatile and rich material platform to explore novel physical phenomena. Here, we report the visualization of an intra-unit cell nematic order concomitant with a Fermi surface distortion at and near the Fermi energy detected via scanning tunneling microscopy and angle-resolved photoemission spectroscopy. These observations serve as clear indicators for the Pomeranchuk instability in α-As.

The crystalline lattice of α-As has a rhombohedral primitive unit cell within the space group $R\bar{3}m$ (No. 166). However, we conveniently adopt the hexagonal (conventional) unit cell representation for our real-space measurements,. The α-As crystal is composed of stacked As bilayers oriented along the *c*-axis (Fig. 1**A**). Within each bilayer, the As atoms arrange themselves in a buckled honeycomb lattice along the *ab*-plane. Mechanical cleavage of the crystal along the *ab*-plane leaves a triangular lattice, as depicted in Fig. 1**B**. We image this triangular lattice (Fig. 1**C**) through scanning tunneling microscopy collected from a freshly cleaved sample at $T = 150$ K. The corresponding Fourier transform image, displayed in Fig. 1**D**, reveals a six-fold rotational symmetry ($C_6$) as indicated by the six-fold symmetric Bragg peak intensities ($Q_1$, $Q_2$, and $Q_3$). The averaged tunneling conductance (d$I$/d$V$) spectrum, which probes the local density of states, obtained from this surface at $T = 150$ K indicates a metallic state. However, upon cooling the sample, we observe the emergence of a stripe-like order. This is clearly visible in the topographic images shown in Figs. 1**F** ($T = 50$ K) and 1**I** ($T = 4$ K). At both temperatures, the Fourier transform images unveil a significantly larger intensity of the $Q_3$ Bragg peak when compared to those associated to $Q_1$ and $Q_2$ (Figs. 1**G** and 1**J**). This change in Bragg peak intensities points to a reduction in the rotational symmetry from $C_6$ to $C_2$, indicative of a nematic state. Intriguingly, the averaged d$I$/d$V$ spectra at these temperatures reveal the renormalization of spectral weight, suggesting a suppressed local density of states precisely at, or very close to the Fermi energy (Figs. 1**H** and 1**K**). This suppression likely results from electronic correlations, which should play a role in the observed nematicity. It is also worth noting that we observed nematic domains exhibiting distinct preferred orientations (Fig. S1), indicating spontaneous symmetry breaking.

Having visualized the nematic order, we now investigate its energy dependence, as summarized in Figs. 2, S2, and S3. Through topographic and spectroscopic imaging at various tip-sample biases, we find that the stripe-like nematic order is prominently present precisely at, and in the immediate vicinity of, the Fermi energy. This is evident in the -5 mV data in Fig. 2**C** and the 0 mV data in Fig. S2, where the $Q_3$ Bragg peak exhibits a higher intensity compared to both $Q_1$ and $Q_2$. This pattern reverses As we move slightly away from the Fermi energy, with $Q_1$ and $Q_2$ showing higher intensities than $Q_3$, as shown in the -30 mV and 30 mV data in Figs. 2**B** and 2**D**. This observation indicates that while the nematic order persists, there is a change in the charge configuration. Above 50 mV or below -50 mV (Fig. S3), the manifestations of nematic order fade, and $Q_1$, $Q_2$, and $Q_3$ exhibit similar intensities, thereby restoring the $C_6$ symmetry of the underlying lattice of the α-As *ab*-plane. Figures 2**A** and 2**E** visually depict the representative $C_6$ symmetric states at -100 mV and 100 mV, respectively. The energy-dependent behavior of the nematic order underscores that nematicity primarily occurs precisely at or near the Fermi energy.

To capture the low-energy electronic response of this nematic state, we conducted quasiparticle interference spectroscopy. We selected a large area on the α-As *ab*-plane, as illustrated in Fig. 3**A**, to perform energy- and spatially-resolved d*I*/d*V* measurements. The resulting d*I*/d*V* maps exhibit real-space patterns that vary with bias voltage (Fig. 3**B**) due to quasiparticle scattering on the surface. By Fourier-transforming the d*I*/d*V* maps, we derive the quasiparticle interference data depicted in Fig. 3**C**. In this data, a ring-shaped signal centered at the scattering $q = 0$ vector corresponds to the Rashba surface state [24, 25]. As the bias voltage increases, the radius of this pattern expands, as anticipated from the band structure of α-As. Additionally, smaller $q$ features are visible, such as a star-shaped pattern with six-fold symmetry observed at 100 mV. This pattern stems from the bulk states [25].

Here, we focus on the ring pattern centered at $q = 0$, where we make a crucial observation. At ±100 mV, the ring displays isotropic behavior, with the scattering vectors $q_1$ and $q_2$ representing scattering vectors along the two high symmetry $\bar{\Gamma} - \bar{M}$ directions of the surface first Brillouin zone (see Fig. 4**A** for the Brillouin zone), with both directions exhibiting similar $q$ vector magnitudes. Strikingly, near the Fermi energy, the ring pattern undergoes a deformation, transforming into an elliptical shape. Specifically, $q_1$ surpasses $q_2$, as is clearly visible in the -12.5 mV, 0 mV, and 25 mV data in Fig. 3**C**. Intriguingly, this deformation is most pronounced precisely at the Fermi energy. To visually represent this discovery, Fig. 3**D** plots the dispersion of the quasiparticle interference pattern along the $q_1$ and $q_2$ directions. From this, we extracted the dispersion of the scattering wavevector of the Rashba surface state (depicted in the lower panel of Fig. 3**E**). Finally, we plot the anisotropy, defined as the ratio of $q_1$ to $q_2$, as a function of the energy (upper panel of Fig. 3**E**). This graphical representation clearly shows that the anisotropy becomes evident only within the -50 mV to 50 mV energy range, peaking precisely at the Fermi energy. This energy window, where the anisotropy is observable, aligns remarkably well with the energy window where the nematic order is observed (Fig. S3). The deformation of the Rashba Fermi surface at and near the Fermi energy, which transforms the circular Fermi surface into an elliptical one, breaking of the $C_6$ rotational symmetry of the (surface) Rashba Fermi surface, echoes the characteristics of the Pomeranchuk instability [11].

Finally, to firmly establish the observation of a deformation of the Rashba Fermi surface, we conducted angle-resolved photoemission spectroscopy measurements. We mapped the Fermi surface at different energies (Fig. 4). Consistent with our scanning tunneling microscopy observations, we found that at the Fermi energy, the Rashba Fermi surface is elliptical. At energies lower than -50 meV, the Rashba Fermi surface appears to be reasonably circular (Fig. 4**B**). By extracting the energy-momentum dispersion along the three high-symmetry $\bar{\Gamma} - \bar{M}$ directions (Fig. 4**C**) and perpendicular to it (Fig. 4**D**) within the surface Brillouin zone, we visualize an anisotropic dispersion that is pronounced at and near the Fermi energy. This observation suggests the breaking of the $C_6$ rotational symmetry in the Rashba Fermi surface. Notably, this observation contrasts with the bulk Fermi surface (Fig. S4), which maintains the $C_3$ symmetry of the bulk Brillouin zone.

Thus, our combined scanning tunneling microscopy/spectroscopy and angle-resolved photoemission spectroscopy study illustrates a distinct deformation of the Fermi surface within the same energy range where the electronic nematicity emerges. Specifically, we observe the transformation of the initially circular Rashba-like Fermi surface into an elliptical shape precisely at and near the Fermi energy, where the electronic nematic order emerges. This observation aligns with the scenario of a Pomeranchuk instability-driven nematic state. To understand the basic phenomenology of the Fermi surface deformation, we have developed a minimal theoretical model (discussed in the Supplementary Materials). We used the time-reversal and surface crystal symmetries to derive an effective $k \cdot p$ non-interacting Hamiltonian describing the electron surface states. The Fermi surface has rotational $C_6$ symmetry due to a combination of $C_3$ lattice and time-reversal symmetries. We show that the electron interactions can potentially drive a phase transition towards a nematic order parameter in the *d*-wave angular momentum channel, which is a two-dimensional irreducible representation in the presence of $C_3$ (or $C_6$) symmetry. By spontaneously choosing a linear representation from this space, the rotational symmetry is broken to $C_2$. The phase transition occurs within a reasonable range of parameters. This suggests that the Fermi liquid state in α-As is indeed susceptible to the Pomeranchuk instability, which is the most likely mechanism for the observed nematicity.

As for the observation of a Pomeranchuk instability on the surface but not in the bulk of α-As, one should bear in mind that the Lindhard function [26], which quantifies the electric field screening by electrons, displays discontinuities only at dimensions below three, exposing the instability of the electron gas or its tendency to order at low dimensions.

We conclude by stating that many observations of electronic nematicity are poorly understood. For example, nematicity has been observed in twisted bilayer graphene, but it remains unclear if it is intrinsic or extrinsic, *i.e.*, strain-induced [27]. In Fe-based superconductors, nematicity is accepted to have an electronic origin likely driven by magnetic fluctuations, although charge fluctuations cannot be discarded [8]. In the case of α-As, there is no evidence for relevant magnetic fluctuations, given that magnetic order is absent. Notice that our structural analysis discards a lattice-driven phase transition (Fig. S5). Therefore, given the experimental evidence at hand, and in contrast to graphene or Fe-based superconductors, we ought to conclude that nematicity in α-As is likely driven by a Pomeranchuk instability, although occurring solely on the surface of this element. This finding is remarkable since Fermi surface instabilities arising from electronic correlations are typically associated with strongly correlated fermionic systems, *e.g.*, in the presence of van Hove singularities. Surface states of topological insulators, on the other hand, are conventionally described via weakly interacting single-particle Hamiltonians. In fact, topological surface states typically do not support strongly interacting broken symmetry states of matter. Therefore, the presence of Pomeranchuk instability driven by interaction on the topological surface state is a surprise and demands developments to current theoretical models. To our knowledge, this is the first evidence of a topological surface state exhibiting strong interaction physics (that breaks symmetry).

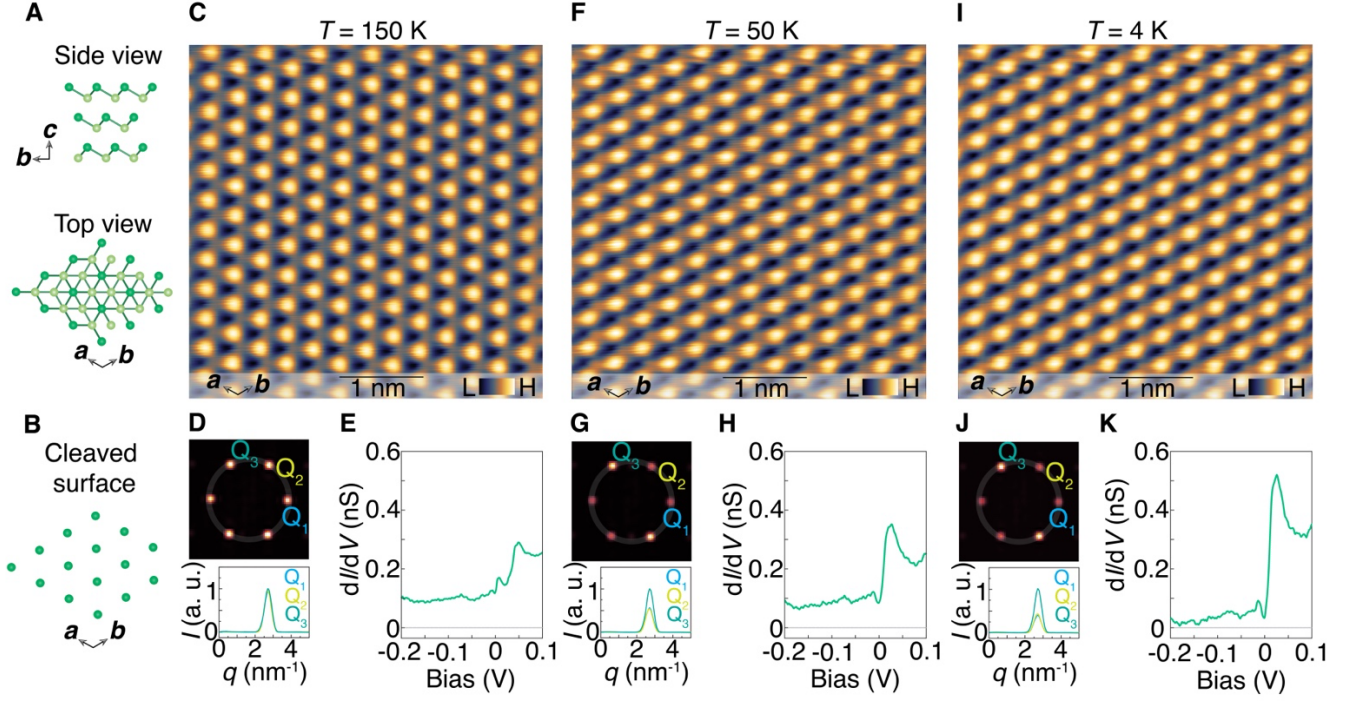

**Fig. 1: Electronic nematicity in α-As at low temperatures. (A)** Upper panel: Side view of the α-As crystallographic structure, consisting of As bilayers stacked along the *c*-axis. Lower panel: Top view of the crystal structure showing the As atoms within the bilayer forming a buckled honeycomb lattice. **(B)** Cleaved *ab*-plane, featuring a triangular lattice. **(C)** Scanning tunneling microscopy topographic image of the α-As *ab*-plane ($V_{gap}$ = -5 mV, $I_t$ = 1 nA), acquired at $T$ = 150 K, exposing its triangular lattice. H and L labels in the color bar denote high and low scale, respectively. **(D)** Upper panel: Fourier Transform of the topographic image in panel **C** revealing three sharp Bragg peaks ($Q_1$, $Q_2$, and $Q_3$) resulting from the triangular lattice. Lower panel: Intensity of all three Bragg peaks, showing similar values for $Q_1$, $Q_2$, and $Q_3$. The intensities are extracted by taking line cuts in the upper panel. **(E)** d$I$/d$V$ spectrum acquired at $T$ = 150 K. **(F)** Topographic image of the α-As *ab*-plane ($V_{gap}$ = -5 mV, $I_t$ = 1 nA), obtained at $T$ = 50 K, visualizing an emerging intra-unit cell stripe-like order, which is absent at $T$ = 150 K (panel **C**). **(G)** Upper panel: Fourier Transform of the topographic image in panel **F**. Lower panel: Intensity of the three Bragg peaks, $Q_1$, $Q_2$, and $Q_3$. At $T$ = 50 K, the intensities of $Q_1$, $Q_2$, and $Q_3$ break the original six-fold rotational symmetry of the α-As *ab*-plane. The intensity values were extracted by taking line cuts in the upper panel. **(H)** d$I$/d$V$ spectrum acquired at $T$ = 50 K. **(I)** Topographic image of the α-As *ab* plane ($V_{gap}$ = -5 mV, $I_t$ = 1 nA), collected at $T$ = 4 K, exposing a pronounced stripe-like order, as seen in panel **F**. **(J)** Upper panel: Fourier Transform of the topographic image in panel **I**. Lower panel: Intensity of the three Bragg peaks, $Q_1$, $Q_2$, and $Q_3$. Akin to the data in panel **G**, the intensities of $Q_1$, $Q_2$, and $Q_3$ break the original six-fold rotational symmetry of the *ab*-plane of α-As. The intensities are extracted by taking line cuts in the upper panel. **(K)** d$I$/d$V$ spectrum acquired at $T$ = 4 K.

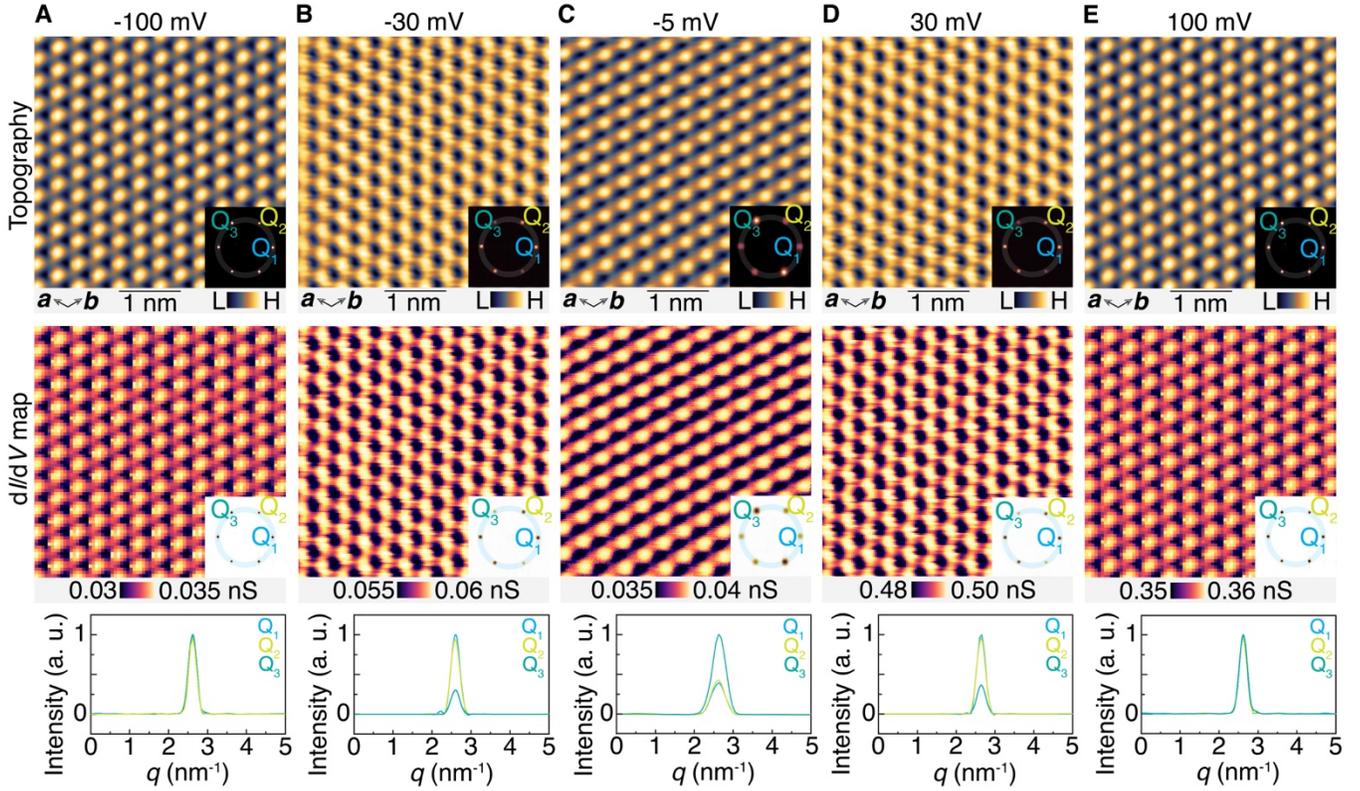

**Fig. 2: Energy dependence of the rotational symmetry-broken order. (A-E)** Topographic images (first row), corresponding d$I$/d$V$ maps (second row), and intensity of the three Bragg peaks, $Q_1$, $Q_2$, and $Q_3$ (third row) acquired at different tip-sample biases, ranging from -100 mV to 100 mV. All the data were collected at $T = 4$ K. Insets in the first and second rows depict the respective Fourier transform images for each panel, indicating the Bragg peaks, $Q_1$, $Q_2$, and $Q_3$. The intensities of $Q_1$, $Q_2$, and $Q_3$ shown in the third-row plots are extracted by taking linecuts in the Fourier transform of the corresponding d$I$/d$V$ maps. Consistent with the Fig. 1 data, topographic and spectroscopic imaging at -5 mV (panel **C**) reveals a stripe-like pattern, which breaks the six-fold rotational symmetry. Data at -30 mV (panel **B**) and 30 mV (panel **D**) exhibit a flip in the intensity of the Bragg peak when compared to -5 mV data, suggesting a redistribution of charges. Nevertheless, the charge distribution still retains a nematic (or two-fold symmetric) order at -30 mV and 30 mV. At -100 mV (panel **A**) and 100 mV (panel **E**), the six-fold rotational symmetry of the α-As *ab*-plane is restored. Thus, the rotational symmetry-broken charge order is only present at or near the Fermi energy. Tunneling junction set-up for topographic images acquired at different bias voltages: $I_t = 1$ nA. Tunneling junction set-up for d$I$/d$V$ maps: $V_{set} = 100$ mV, $I_{set} = 1$ nA, $V_{mod} = 3$ mV.

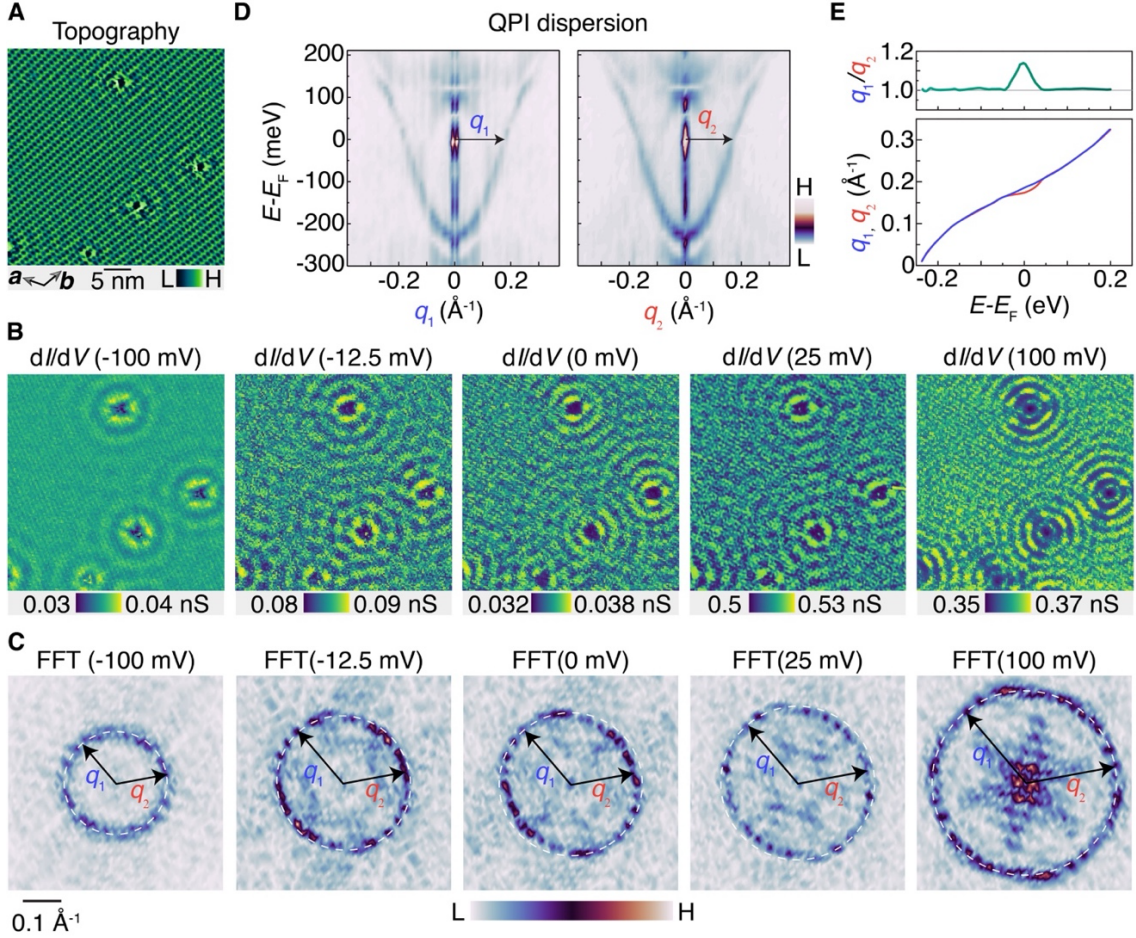

**Fig. 3: Visualization of the deformation of the Rashba Fermi surface and concomitant $C_6$ symmetry breaking through quasiparticle interference.** **(A)** Large-scale topography of the α-As *ab* plane revealing several scattering sources. **(B)** Spectroscopic maps and **(C)** their corresponding Fourier transforms for various tip-sample biases, ranging from -100 mV to 100 mV. Clear quasiparticle interference patterns are observed at all the bias voltages. At -100 mV, a circular pattern centered at (0,0) corresponds to the Rashba surface state of α-As. The radius of this circular pattern expands as the bias voltage increases. Anisotropic quasiparticle interference patterns emerge at -12.5 mV, 0 mV (Fermi energy), and 25 mV, with the most pronounced anisotropy observed at 0 mV. However, at 100 mV, the pattern returns to a circular shape. White dashed circles in each panel in **C** serve as guides to the eyes to expose the anisotropy in the quasiparticle interference of the Rashba Fermi surface. The momentum vectors $q_1$ and $q_2$ denote the scattering branch of the Rashba surface state along the two high symmetry $\bar{\Gamma} - \bar{M}$ directions of the surface first Brillouin zone. **(D)** Quasiparticle interference spectra acquired along $q_1$ (left) and $q_2$ (right) directions, showing multiple interference branches. The parabolic-shaped branch, originating from $\sim -(230 \pm 12.5)$ meV, represents the Rashba surface state. The directions of $q_1$ and $q_2$ are indicated in panel **C**. **(E)** Lower panel: Dispersion of the scattering branch from the Rashba surface state along the $q_1$ and $q_2$ directions, as extracted from the data in **D**. The quasiparticle interference pattern exhibits a finite bandwidth. The magnitude of $q_1$ and $q_2$ are determined at each energy by the median value of the bandwidth. Upper panel: Anisotropy of the Rashba surface state scattering branch, defined as the ratio between the magnitudes of $q_1$ and $q_2$, plotted as a function of the energy. The anisotropy is most pronounced at and near the Fermi energy. All the data were collected at $T = 4$ K using the following tunneling junction set-up for d$I$/d$V$ maps: $V_{set}$ = 200 mV, $I_{set}$ = 1 nA, $V_{mod}$ = 5 mV.

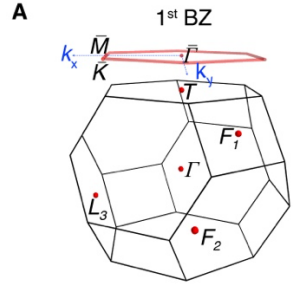
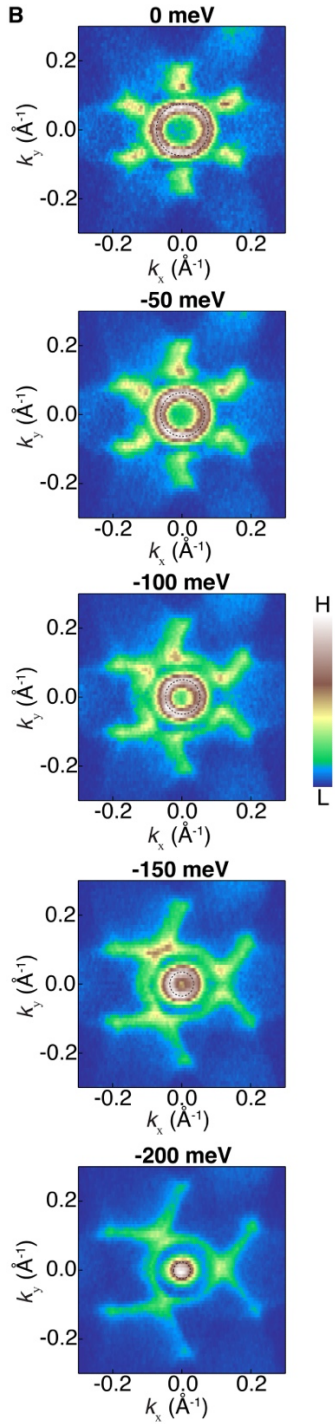
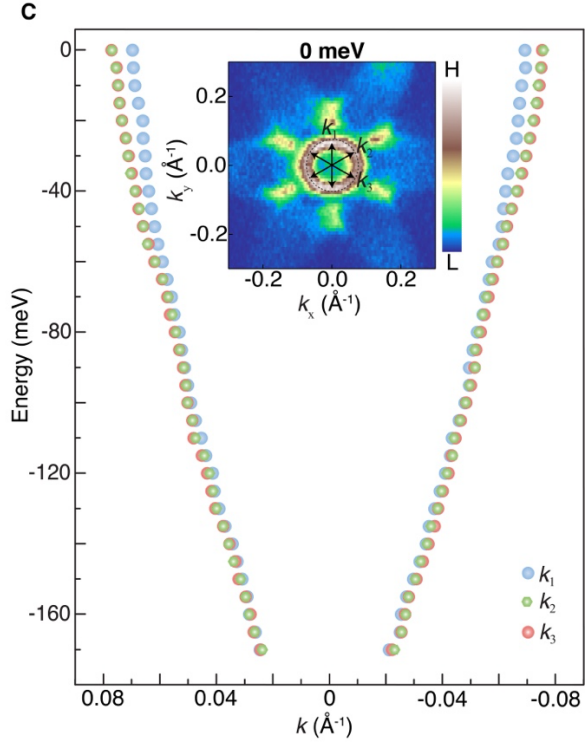
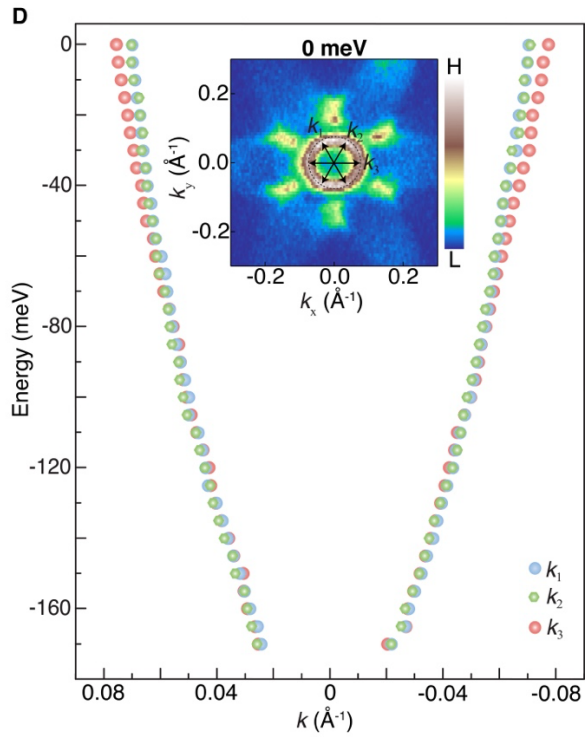

**Fig. 4: Observation of Rashba Fermi surface deformation and resulting C$_6$ symmetry breaking through angle-resolved photoemission spectroscopy.** **(A)** First Brillouin zone and the projected surface along the (111) $k$-direction. **(B)** Fermi surface map taken on the As (111) surface. The map is obtained using 84 eV photon energy with linear-horizontally polarized light. In addition to the three-dimensional bulk state, the photoemission data reveal ring-shaped Fermi pockets centered at $k = 0$, stemming from the Rashba surface state. The radius of this ring-shaped pattern shrinks as the energy is lowered. The ring exhibits anisotropy at the Fermi energy, with the anisotropy becoming less prominent away from the Fermi energy. Black dashed circles in each panel serve as guides to visualize the anisotropy of the Rashba Fermi surface. **(C)** Energy-momentum dispersion of the Rashba surface state along the $\bar{\Gamma} - \bar{M}$ directions, acquired along $k_1$, $k_2$, and $k_3$ wavevectors as indicated in the inset depicting the Fermi surface map. **(D)** Energy-momentum dispersion of the Rashba surface state perpendicular to $\bar{\Gamma} - \bar{M}$ directions, acquired along $k_1$, $k_2$, and $k_3$ as shown in the inset. The bands display a finite bandwidth. The magnitudes of $k_1$, $k_2$, and $k_3$ in panels **C** and **D** are determined at each energy by the median value of the bandwidth. These bands demonstrate anisotropic energy-momentum dispersion at and near the Fermi energy (akin to the quasiparticle interference data in Fig. 3). All the data were obtained at $T = 10$ K.

**Acknowledgement:**
M.Z.H. group acknowledges primary support from the US Department of Energy (DOE), Office of Science, National Quantum Information Science Research Centers, Quantum Science Center (at ORNL) and Princeton University; STM Instrumentation support from the Gordon and Betty Moore Foundation (GBMF9461) and the theory work; and support from the US DOE under the Basic Energy Sciences programme (grant number DOE/BES DE-FG-02-05ER46200) for the theory and sample characterization work including ARPES. Z.M. and W.Z. acknowledge support from the National Natural Science Foundation of China (Grant No. 62150410438), the Beihang Hefei Innovation Research Institute (project no. BHKX-19-02) for growth and beamline 13U of the National Synchrotron Radiation Laboratory (NSRL) for the ARPES experiments. R.I. acknowledges support from the Foundation for Polish Science through the international research agendas program co-financed by the European union within the smart growth operational program and the National Science Foundation under Grant No. OIA-2229498, University of Alabama at Birmingham (UAB) internal startup funds, and UAB Faculty Development Grant Program, Office of the Provost. L.B. is supported by DOE-BES through award DE-SC0002613. The National High Magnetic Field Laboratory (NHMFL) acknowledges support from the US-NSF Cooperative agreement Grant number DMR- 2128556. and the state of Florida. T.N. acknowledges support from the Swiss National Science Foundation through a Consolidator Grant (iTQC, TMCG-2_213805).